# All-optical magneto-thermo-elastic skyrmion motion

Serban Lepadatu[1]

[1]*Jeremiah Horrocks Institute for Mathematics, Physics and Astronomy, University of Central Lancashire, Preston PR1 2HE, U.K.*

**Abstract**

It is predicted magnetic skyrmions can be controllably moved on surfaces using a focused laser beam. Here an absorbed power of the order 1 mW, focused to a spot-size of the order 1 µm, results in a local temperature increase of around 50 K, and a local perpendicular strain of the order $10^{-3}$ due to the thermo-elastic effect. For positive magneto-elastic coupling this generates a strong attractive force on skyrmions due to the magneto-elastic effect. The resultant motion is dependent on forces due to i) gradients in the local strain-induced magnetic anisotropy, ii) gradients in the effective anisotropy due to local temperature gradients, and magnetic parameters temperature dependences, and iii) Magnus effect acting on objects with non-zero topological number. Using dynamical magneto-thermo-elastic modelling, it is predicted skyrmions can be moved with significant velocities (up to 80 m/s shown), both for ferromagnetic and antiferromagnetic skyrmions, even in the presence of surface roughness. This mechanism of controllably moving single skyrmions in any direction, as well as addressing multiple skyrmions in a lattice, offers a new approach to constructing and studying skyrmionic devices with all-optical control.

[*]SLepadatu@uclan.ac.uk

# I. Introduction

Magnetic skyrmions are disk-like topological objects in magnetization textures [1], which can be stabilized using the Dzyaloshinskii-Moriya interaction (DMI) [2,3] in materials lacking inversion symmetry, either in the bulk [4], but also at room temperature in layers with interfacial DMI [5]. Applications have been proposed for skyrmionic devices, including artificial synapses for neuromorphic computing [6], neural networks [7], skyrmionic interconnects [8], probabilistic computing [9], and magnetic memory [10]. Moving skyrmions is typically achieved using electrical contacts, relying on spin torques due to charge and spin currents, including spin-orbit torques [11-13], as well as bulk and interfacial spin transfer torques [14,15]. Other methods include use of surface acoustic waves (SAW) generated using interdigitated transducers [16], magnetic field gradients generated using a magnetic force microscopy (MFM) probe [17], thermal gradients generated using on-chip heaters [18], and even the combination of temperature gradients with an externally generated strain has been proposed [19]. Apart from the use of MFM probes, all these methods lack the ability to control motion in two dimensions on magnetic surfaces.

It is known that skyrmions may be created and annihilated optically using laser pulses [20-22]. For all-optical control, a missing element is the ability to move magnetic skyrmions using laser beams. For magnetoelectric materials it has been proposed that multiferroic skyrmions may be manipulated using the electric field from a laser beam [23], and an inhomogeneous electric torque could be used to drive skyrmions in multiferroic or magnetic insulators [24]. Here it is proposed that laser beams may also be used to manipulate ferromagnetic and antiferromagnetic skyrmions, relying on the magneto-elastic (ME) effect due to thermo-elastic lattice expansion, as well as local temperature gradients. The effect of temperature gradients on skyrmions has been discussed [25]. Anisotropy gradients in particular have been shown to generate skyrmion motion [26], with nanopatterning of the anisotropy landscape used to guide current-induced skyrmion motion [27], and entropic forces due to temperature gradients generate skyrmion Seebeck and Nernst effects [28]. Apart from forces arising directly from temperature gradients, an important effect which has not been discussed so far in the context of skyrmion motion, is thermo-elastic lattice expansion resulting in a strain-induced magnetic anisotropy modification. Thermo-elastic lattice expansion can be important, for example it can result in an enhanced magnonic spin Seebeck effect [29], and can even induce magnetization dynamics and spin currents in magnetic insulators [30]. Moreover, it is inextricably linked to temperature gradients, resulting in a magneto-thermo-elastic coupling effect, and a key message in this work is that an additional force due to lattice expansion arises on skyrmions, which must be taken into consideration.

From a practical perspective for potential applications, heat-assisted magnetic recording [31] already uses laser beams to control the magnetic anisotropy, relying on rapid spot heating approaching the Curie temperature. For all-optical magnetic skyrmion motion, it is shown here spot heating up to 350 K is sufficient. Moreover, it is possible to generate and control on-chip laser beams, for example using a microelectromechanical systems-based (MEMS) steering system, achieving 3.3 μm diameter focused beam with steering speeds of 10 m/s over a travel distance of 10 μm [32], or using MEMS-controllable vertical cavity surface emitting laser (VCSEL) microlens arrays [33], with more recent efforts demonstrating MEMS-VCSEL multi-MHz rastering capability [34].

## II. Magneto-Thermo-Elastic Model

The model is based on three coupled dynamics equations for magnetization, temperature, and elasticity properties:

$$\frac{\partial \mathbf{m}}{\partial t} = -\gamma \mathbf{m} \times \mathbf{H}_{eff} + \alpha \mathbf{m} \times \frac{\partial \mathbf{m}}{\partial t}$$
$$C\rho \frac{\partial T}{\partial t} = K \nabla^2 T + Q \qquad (1)$$
$$\rho \frac{\partial v_p}{\partial t} = \sum_{q=x,y,z} \frac{\partial \sigma_{pq}}{\partial q} - \eta v_p, \quad (p = x, y, z)$$

The first equation is the Landau-Lifshitz-Gilbert (LLG) equation, where **m** is the magnetization direction, $\gamma$ is the gyromagnetic ratio, and $\alpha$ is the Gilbert damping factor [35]. Here $\mathbf{H}_{eff}$ is a total effective field which includes: i) demagnetizing field [36], ii) uniaxial magneto-crystalline anisotropy, $\mathbf{H} = 2K_U(\mathbf{m}.\mathbf{e}_A)\mathbf{e}_A / \mu_0 M_S$, where $K_U$ is the uniaxial anisotropy constant, $M_S$ is the saturation magnetization, and $\mathbf{e}_A = \hat{\mathbf{z}}$ is the easy axis direction perpendicular to the plane, iii) direct exchange interaction, $\mathbf{H} = 2A\nabla^2 \mathbf{m} / \mu_0 M_S$, where $A$ is the exchange stiffness constant, iv) interfacial DMI, $\mathbf{H} = 2D((\nabla.\mathbf{m})\hat{\mathbf{z}} - \nabla m_z) / \mu_0 M_S$, where $D$ is the DMI constant, v) external field, and vi) ME interaction. Thus the LLG equation is coupled to the heat and elastodynamics equations through the temperature dependence of magnetic parameters, and through the ME interaction respectively. The latter is due to spin-orbit interaction and has the energy density expression (ME field obtained as $\mathbf{H} = -(\partial E_{ME} / \partial \mathbf{m}) / \mu_0 M_S$):

$$E_{ME} = B_1[m_x^2 \varepsilon_{xx} + m_y^2 \varepsilon_{yy} + m_z^2 \varepsilon_{zz}] + 2B_2[m_x m_y \varepsilon_{xy} + m_x m_z \varepsilon_{xz} + m_y m_z \varepsilon_{yz}] \quad (2)$$

Here $B_1$ and $B_2$ are ME coupling coefficients, and $\boldsymbol{\varepsilon}$ is the symmetric strain tensor with indicated linear and shear strain components. The second expression in Equation (1) is the heat equation, where $C$ is the specific heat capacity, $\rho$ is the mass density, $K$ is the thermal conductivity, $T$ is the temperature, and $Q$ is a heat source. Here the heat source is due to a laser beam spot with Gaussian profile $Q = Q_0 \exp(-2r^2 / r_0^2)$, where $r$ is the radial distance from the spot centre, $Q_0$ is the laser beam power density, and $2r_0$ is the diffraction-limited spot-size. On the one hand the temperature increase gives rise to a temperature dependence of magnetic parameters, namely the magnetization length is scaled according to the Curie-Weiss law [37], $m_e(T) = B(m_e 3T_C / T)$, where $B$ is the Langevin function and $T_C$ is the Curie temperature. The uniaxial magneto-crystalline anisotropy constant is scaled as $m_e^3$ [38], the exchange stiffness and DMI constant are scaled as $m_e^2$ [39,40]. Finally, the damping constant is scaled as $(1 - T / 3T_C)$ [37]. On the other hand the temperature increase gives rise to a mechanical stress in the elastodynamics equation, due to the thermo-elastic effect. This is the third expression in Equation (1), written in the velocity-stress representation [41], where $\mathbf{v}$ is the velocity, $\boldsymbol{\sigma}$ is the symmetric stress tensor, and $\eta$ is a mechanical damping factor resulting in energy dissipation. This is solved using the finite-difference time-domain (FDTD) scheme, with implementation details for the 3D elastodynamics equation in multi-layered structures given in Appendix A. The stress and strain for a cubic crystal are related using:

$$\sigma_{pp} = c_{11}\varepsilon_{pp} + 2c_{12}(\varepsilon_{qq} + \varepsilon_{rr}) - (c_{11} + 2c_{12})\alpha_T(T - T_0) - B_1 m_p^2, \quad (p,q,r = x,y,z, \; p \neq q \neq r)$$
$$\sigma_{pq} = c_{44}\varepsilon_{pq} - 2B_2 m_p m_q, \quad (p,q = x,y,z, \; p \neq q) \quad (3)$$

Here $c_{11}$, $c_{12}$, $c_{44}$ are elastic stiffness coefficients (taken as $c_{11}$ = 300 GN/m$^2$, $c_{12}$ = 200 GN/m$^2$, and $c_{44}$ = 50 GN/m$^2$ [42]), $\alpha_T$ is the linear thermal expansion coefficient, and in this work $T_0$ is the ambient temperature taken as 300 K. The terms involving $B_1$ and $B_2$ in Equation (3) give the magnetostriction contribution. For use in Equation (2), the strain is obtained by inverting Equation (3). Note, the inverse thermo-elastic effect (change in temperature due to lattice strains) is not considered since it is negligible for the problem studied here.

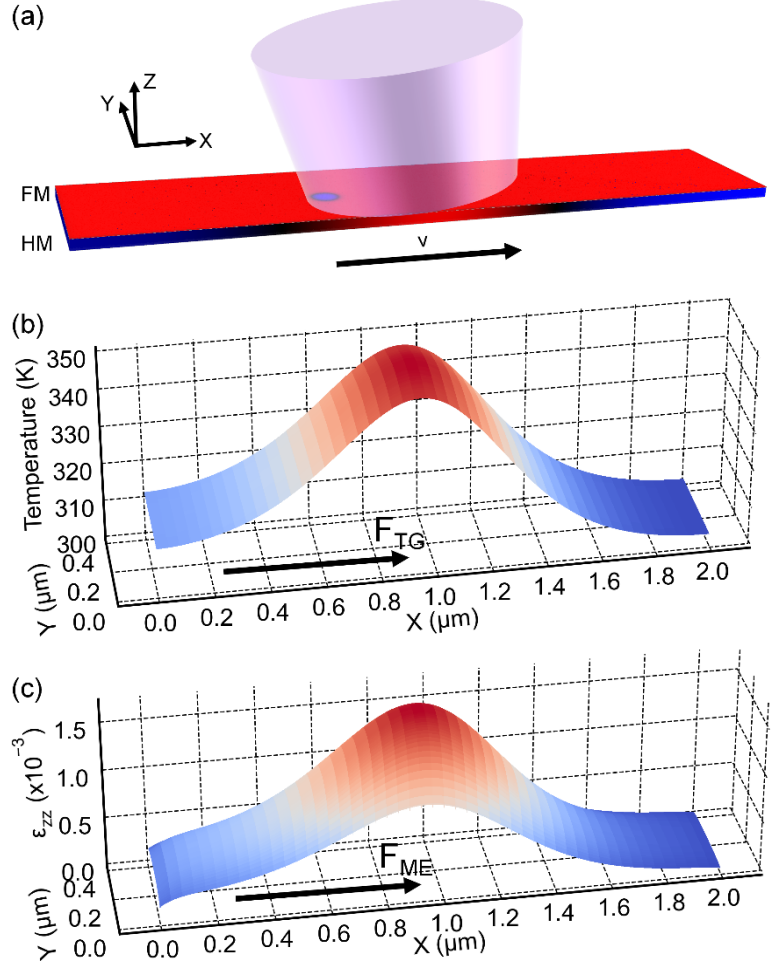

**Figure 1** – Skyrmion motion generated by scanning a focused laser beam, due to magneto-thermo-elastic coupling in a magnetic track. (a) Magnetic track showing a skyrmion being moved using a focused laser beam (1 mW absorbed power) and with a set velocity along the track (v = 20 m/s). In the FM layer, red indicates magnetization out of the plane, and blue into the plane. (b) Computed dynamic temperature profile. A force towards the centre of the laser spot arises directly due to temperature gradients, $F_{TG}$. (c) Computed z-direction dynamic strain profile. A strain-induced graduated magnetic anisotropy is generated, which for positive ME coupling results in an additional force towards the centre of the laser spot, $F_{ME}$.

Figure 1 shows the effect of a laser beam with spot-size 0.8 μm (e.g. possible using a laser wavelength of 442 nm [43]) and $Q_0$ = 2.25×10$^{18}$ W/m$^3$. Energy is absorbed in the FM layer of thickness $t_{FM}$ = 2 nm, and integrating the laser spot Gaussian profile gives the total power $P = Q_0 t_{FM} r_0^2 \pi (1 - e^{-2})/2 \cong 1$ mW. The actual laser power required to produce this depends on the material-specific laser absorption coefficient, thus here we simply state the total absorbed power as used in computations. The structure used is a FM/HM bilayer track hosting Néel skyrmions (e.g. Co/Pt, where $A = 10$ pJ/m, and $D = -1$ mJ/m² [12], although the general physical picture presented here is not restricted to these materials), with FM thickness $t_{FM} = 2$ nm, HM thickness $t_{HM} = 30$ nm, width 0.4 μm, and length 2 μm. Further computational details are given in Appendix A. The laser beam is scanned along the track, through the middle, with a set velocity (20 m/s used here) as indicated in Figure 1(a), resulting in skyrmion motion due to attractive forces towards the centre of the laser spot. Figure 1(b) shows the dynamic temperature profile half-way through the scan, with the maximum temperature increase of ~50 K reached at the centre of the laser spot, decreasing either side. The trailing side has a higher temperature compared to the leading side due to finite heat dissipation time (heat is transferred to the surroundings according to Newton's law of cooling, implemented using Robin boundary conditions [44]). A substrate is not included here for simplicity, although inclusion of a substrate material would result in a slightly larger temperature increase for the same heat source power density [44]. The temperature profile gives rise to an effective anisotropy gradient since magnetic parameters have a temperature dependence, where $K_{eff}(T) = m_e^2 (K_U m_e - \mu_0 M_S^2 / 2)$ is the effective anisotropy, decreasing towards the centre of the laser spot. Here $K_U = 380$ kJ/m³, $M_S = 600$ kA/m [12], and the Curie temperature is taken as 700 K. Anisotropy gradients result in a force on skyrmions [26], and in this case we have a force acting towards the centre of the laser spot, $F_{TG}$, as indicated in Figure 1(b). This is further enhanced due to the increase in skyrmion diameter with decreasing anisotropy, which is known to accelerate skyrmion motion [26]. It should also be noted that the decrease in $A(T)$ and $D(T)$ with temperature has a net opposite effect here (also see Ref. [25]), as the net increase in skyrmion diameter with temperature is reduced. On the other hand, the temperature increase results in sample strains due to the thermo-elastic effect, with the $\varepsilon_{zz}$ strain component shown in Figure 1(c), where the bottom face of the HM layer is mechanically fixed (e.g. as would result when in contact with a stiff substrate). Here $\alpha_T = 10^{-5}$ K⁻¹ is taken as typical order of magnitude linear thermal expansion coefficient in thin metallic films (e.g. for ultrathin Co/Pt bilayers see Ref. [45]). The resulting strain is of the order $10^{-3}$, which generates a significant ME effect. The ME coupling energy density is shown in Equation (2), proportional to the ME coupling coefficient $B_1 = 10$ MJ/m³ (e.g. this value was used to explain the effect of SAWs on skyrmion creation in Co/Pt/Ir tracks [16]). Here we also set $B_2 = B_1$, although the ME effect on skyrmions due to shear strains is negligible here. With a positive ME coupling, a tensile strain as in Figure 1(c) induces a hard magnetic axis, and easy plane perpendicular to the strain. Thus the ME effect is an important additional source of magnetic anisotropy, resulting in an anisotropy gradient and an additional force on skyrmions, $F_{ME}$, which for positive ME coupling also acts towards the centre of the laser spot.

## III. Isolated Skyrmion Motion in Magnetic Tracks

Skyrmion motion in magnetic tracks is now investigated in more detail. A useful model to understand skyrmion motion is the Thiele equation, derived from the LLG equation under simplifying assumptions [46], and extended more recently to include acceleration effects in a linear anisotropy gradient [26]:

$$\mathbf{G} \times \mathbf{v} + \alpha \mathbf{D} \mathbf{v} = -\nabla U \tag{4}$$

Here $\mathbf{G} = \mp 2\pi Q \hat{z}$ is the gyrovector, where $Q = \int (\partial \mathbf{m}/\partial x \times \partial \mathbf{m}/\partial y) dxdy / 4\pi$ is the topological number with the integral taken over the magnetic surface, and for ferromagnetic skyrmions we have $Q = \pm 1$. In Equation (4) $\mathbf{v}$ is the skyrmion velocity, $\alpha \mathbf{D} \mathbf{v}$ is a dissipative term proportional to the Gilbert damping parameter $\alpha$ (here we set $\alpha = 0.1$ [47]), where $\mathbf{D}$ is the dissipative tensor dependent on the skyrmion structure, and $\mathbf{F} = -\nabla E$ is the driving force obtained as the gradient of the potential energy. Thus, for objects with non-zero topological number, the velocity is not along the driving force direction, but acquires an orthogonal component reminiscent of the Magnus effect. As discussed above, the driving forces are $F_{TG}$ and $F_{ME}$, obtained from the gradients of the effective anisotropy and strain-induced anisotropy respectively. These are computed in the Supplemental Material as a function of power, e.g. for $P = 1$ mW, we have 23.8 GN/m$^3$ and 28.4 GN/m$^3$ maximum gradient magnitudes for $F_{TG}$ and $F_{ME}$ respectively. Thus it is expected the force due to ME coupling has a stronger effect than that due to thermal gradients in the case investigated here. Although this depends on the particular material parameters, the key message is the ME effect, as a result of thermo-elastic lattice expansion, has an important contribution, and it is not sufficient to consider only the force due to temperature gradients. Examples of computed skyrmion paths for the track considered above are shown in Figure 2, obtained for $P = 1$ mW and laser spot velocity of 5 m/s, where the laser spot starts at $x = 0.3$ μm and terminates at $x = 1.7$ μm. To see the separate effects of $F_{TG}$ and $F_{ME}$, computations are performed as: i) ME coupling disabled, thus $F_{ME}$ becomes zero, and ii) ME coupling enabled, however the temperature dependence of magnetic parameters is kept constant for $T > 300$ K, and thus $F_{TG}$ becomes zero.

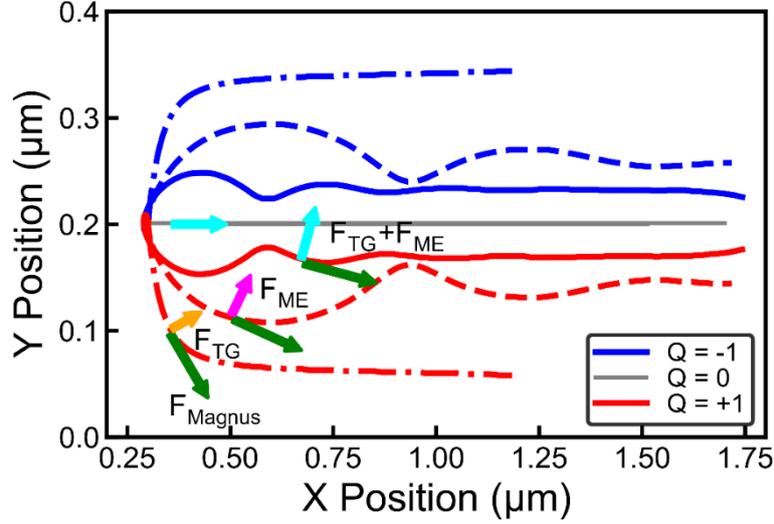

**Figure 2** – Computed skyrmion paths for 1 mW absorbed power, with laser spot scanned at 5 m/s through the centre of a track 400 nm wide and 2 μm long, shown for ferromagnetic skyrmions ($Q = \pm 1$) and antiferromagnetic skyrmion ($Q = 0$). The dash-dot lines show the skyrmion motion without the ME effect, with the resultant motion dependent on the force arising from temperature gradients, $F_{TG}$, and the effective force due to the Magnus effect, $F_{Magnus}$. The dashed lines show the skyrmion motion due to the ME effect, $F_{ME}$, with magnetic parameter temperature dependences disabled. The solid lines show the combined effect, noting that for $Q = 0$ there is no net Magnus force.

These two separate cases are shown in Figure 2 as i) dashed-dot and ii) dashed lines respectively, where an out-of-plane magnetic field of 10 kA/m is used to stabilize a Néel skyrmion with 35 nm starting diameter. The Magnus effect is significant, resulting in a skyrmion velocity component perpendicular to the anisotropy gradient, with sign dependent on the topological number as expected. The computation is also repeated for antiferromagnetic skyrmions ($Q = 0$), using a 2-sublattice LLG equation [48], and as expected the Magnus effect vanishes, with the skyrmion following the laser spot in a straight line. For $Q = 0$ no out-of-plane magnetic field is used, however $D = -2.5$ mJ/m$^2$ in order to stabilize an antiferromagnetic skyrmion also with 35 nm starting diameter; all other magnetic parameters are the same, applied equally to the two sub-lattices. It is also observed the effect of $F_{ME}$ is stronger than that of $F_{TG}$ as expected – indeed, with $F_{TG}$ alone the skyrmion is not displaced for the full laser spot travel distance, but loses tracking half-way through the scan. The full, combined effect, is shown using solid lines in Figure 2. It is observed the skyrmion undergoes an incomplete *bouncing orbit* motion, before settling into an off-centre dynamic equilibrium position. Further information is obtained by investigating the skyrmion velocity as a function of time, shown in Figure 3. For $|Q| = 1$ it is observed the skyrmion undergoes large, damped velocity oscillations for the first 150 ns, before settling into the steady state 5 m/s forward velocity. For the same power of 1 mW, for $Q = 0$ the skyrmion undergoes an initial acceleration in the first 50 ns, before settling into the steady-state velocity of 5 m/s.

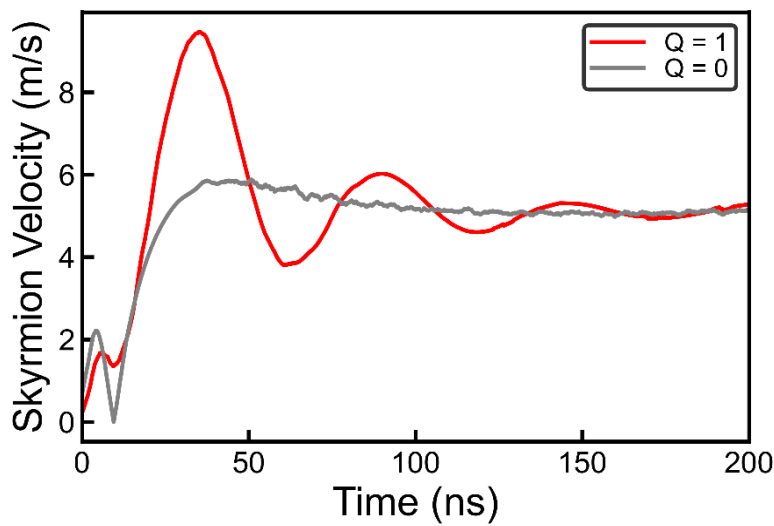

**Figure 3** – Computed skyrmion velocity in response to a laser beam scanned at 5 m/s with 1 mW absorbed power, for ferromagnetic ($|Q| = 1$), and antiferromagnetic skyrmions ($Q = 0$), showing the velocity magnitude. The steady-state target velocity of 5 m/s is reached following an initial damped oscillatory response, particularly pronounced for $Q \neq 0$ where the Magnus force contributes.

Before investigating the effect of varying power and beam velocity, it is important to discuss the effect due to the sign of ME coupling coefficient, $B_1$. Both negative and positive values are possible [49]. Positive ME coupling coefficients have been obtained in skyrmion-supporting MnSi [50-52], whilst positive effective ME coupling in ultrathin Fe films was measured due to different contributions of bulk and surface ME contributions [53]. The ME coupling also depends on the lattice parameter [54], and large positive ME coupling in thin Fe films was measured [55], which was found to change sign and become negative for thicker Fe films. For thin Ni films, positive ME coupling was also measured, with sign dependent on film thickness [56]. For hcp Co large positive values exceeding 10 MJ/m$^3$ have also been measured [57]. Moreover, a value of $B_1 = 10$ MJ/m$^3$ was used to explain the effect of SAWs on magnetic skyrmions in Pt/Co/Ir tracks [16]. If $B_1$ is negative the sign of $F_{ME}$ is reversed, and thus opposes $F_{TG}$. In this case a tensile strain induces an easy axis, and a hard plane perpendicular to the strain axis, with the strain-induced magnetic anisotropy gradient of opposite sign. As shown in the Supplemental Material, indeed the effect of $F_{ME}$ alone for $B_1 = -10$ MJ/m$^3$ is to cause skyrmion motion away from the laser spot centre, although the effect is reduced since the skyrmion diameter is also reduced as the total anisotropy increases. The dissipative term in the Thiele equation, $\alpha \mathbf{D v}$, is dependent on the skyrmion diameter, resulting in smaller velocities for smaller diameters. The combination of $F_{ME}$ and $F_{TG}$ nearly cancels out due to opposite directions on the one hand, with the effect on skyrmion motion also reduced due to reduction in skyrmion diameter. Conversely, when the ME coupling is positive, the combined effect of $F_{ME}$ and $F_{TG}$ is enhanced as the skyrmion diameter increases due to decreasing total anisotropy – the skyrmion diameter in this case, as a function of power, is also shown in the Supplemental Material, which reaches ~120 nm at 1 mW power.

Increasing the power results in stronger forces, as the maximum temperature, perpendicular strain, and therefore anisotropy gradient magnitudes increase (see Supplemental Material for details). Thus it is expected that larger powers allow the skyrmion to be moved with larger velocities. This is shown in Figure 4 both for ferromagnetic ($|Q| = 1$), and antiferromagnetic skyrmions ($Q = 0$). Here the skyrmion path is computed as a function of power and laser beam velocity, with the total skyrmion displacement shown normalized to the laser spot travel distance. If the power is insufficient for a given beam velocity, then skyrmion tracking is lost, which defines region IIb in Figure 4. The boundary delimiting this region is shown, obtained by fitting a logistic function to the normalized skyrmion displacement, $d_{Sk} = 1/(1 + e^{-(P-P_0)/w})$ with transition region width $w$ and threshold power $P_0$. Above this threshold the skyrmion is displaced without losing track of the laser spot, which defines region I. At higher powers it is important to make a distinction however, since the local reduction in total anisotropy results in unstable skyrmions, which are dynamically distorted into *skyrmion strings* (see Supplemental Material for an example), defining region IIa. Comparing thresholds for ferromagnetic and antiferromagnetic skyrmions, it is seen that for $Q = 0$ larger powers are required to fully displace the skyrmion, and also the boundary separating regions I and IIa is shifted up. This is largely because the demagnetizing field is negligible, resulting in a weaker dependence of skyrmion diameter on temperature (see Supplemental Material).

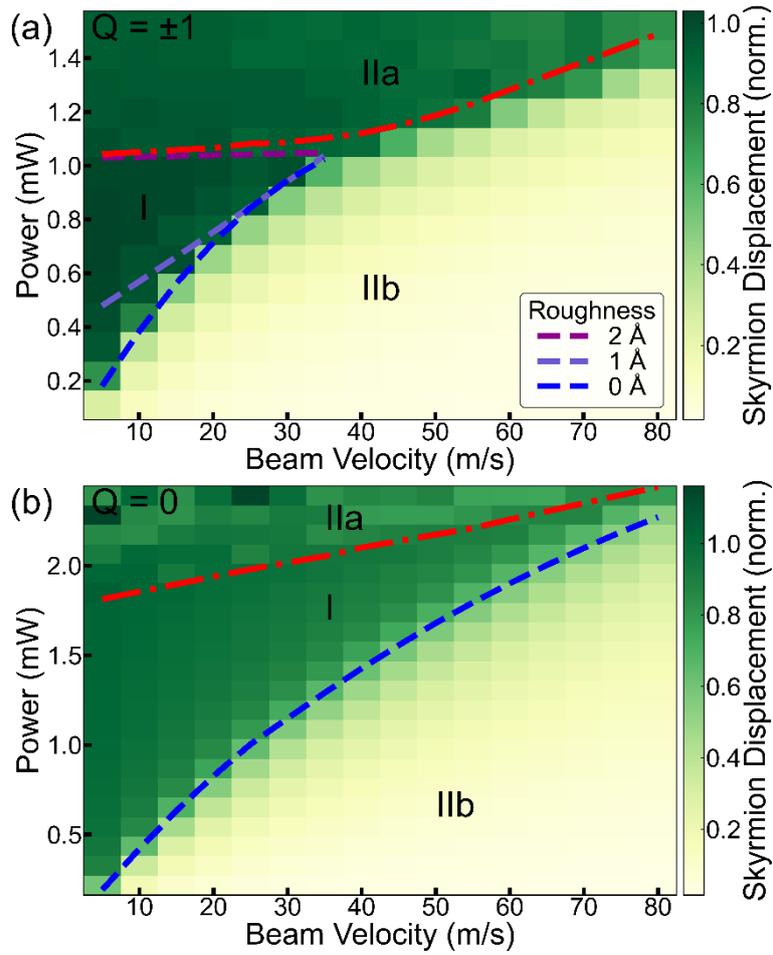

**Figure 4** – Skyrmion displacement (normalized to laser spot travel distance) as a function of power and laser beam velocity, for a) ferromagnetic skyrmion, and b) antiferromagnetic skyrmion. In region IIb the power is insufficient to fully displace the skyrmion, characterized by loss of tracking. In region IIa, at larger powers, the skyrmion is dynamically distorted into a *skyrmion string*, delineated by the dash-dot line. In region I the skyrmion is displaced fully without loss of tracking and without distortion. Boundaries between regions I and IIb are indicated by dashed lines, and for the ferromagnetic skyrmion are computed with different amplitudes of surface roughness, 0 Å (perfect surface), 1 Å, and 2 Å.

Finally, the effect of landscape disorder is briefly addressed. For current-induced skyrmion motion it is known that disorder results in increased threshold current densities, reduced velocities and a variation of the skyrmion Hall angle with velocity [58,59,60]. Here landscape disorder is introduced in the form of surface roughness [60] using an effective field model developed previously [61], considering cases with 1 Å and 2 Å roughness amplitude, and in-plane correlation length of 40 nm. As shown in Figure 4(a), increasing the roughness amplitude results in a shift of the threshold boundary separating regions IIb and I to higher powers. This is expected since larger forces are required to depin skyrmions from pinning potentials formed by the landscape disorder. These results demonstrate that all-optical skyrmion motion, using the magneto-thermo-elastic coupling mechanism, is feasible even when sample imperfections are present, although as the case with 2 Å roughness amplitude shows, excessive imperfections can be problematic for achieving distortion-free skyrmion displacement.

## IV. Directed Skyrmion Motion

One important advantage the all-optical skyrmion motion mechanism presented here has over conventional current-induced skyrmion motion, and even SAW-induced skyrmion motion, is the ability to controllably move skyrmions on magnetic surfaces in any direction. An example of directed skyrmion motion is shown in Figure 5, where a magnetic surface of 2 μm × 2 μm is used, and the laser spot is rotated about the centre with a 700 nm radial distance.

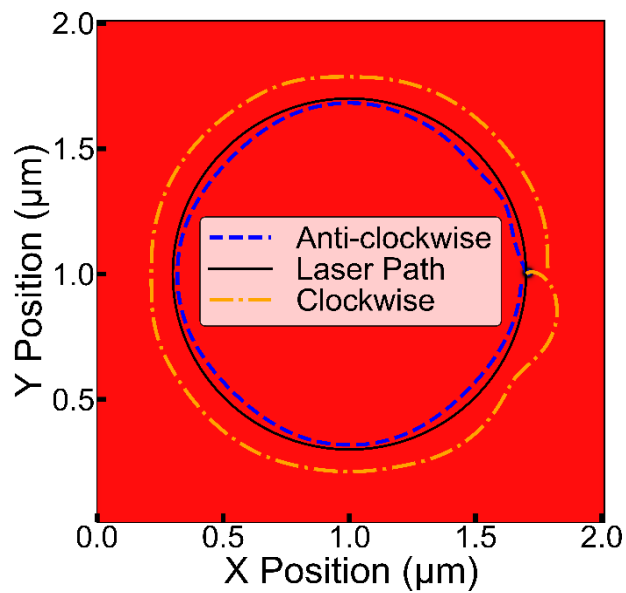

**Figure 5** – Directed skyrmion ($Q = -1$) motion due to a circular laser path, with 1 mW power and 5 m/s tangential velocity (period of 880 ns). The plotted skyrmion paths are computed for clockwise and anti-clockwise laser path directions.

The starting position for the skyrmion and laser spot are $x = 1.7$ μm, $y = 1$ μm as indicated in Figure 5, and two cases are investigated for clock-wise and anti-clockwise laser spot paths. Using a power of 1 mW and linear velocity of 5 m/s, in both cases the skyrmion is observed to track the laser spot fully. Due to the Magnus effect the skyrmion paths differ for the clockwise and anti-clockwise directions, with the clockwise direction causing the skyrmion path to stabilize on the outside of the laser path circle. This results in a larger distance between the skyrmion and laser spot centre compared to the anti-clockwise direction, since the laser spot path is always curving away from the skyrmion. Unlike the case of magnetic tracks, it is expected that the two movement directions now have different power thresholds for any given beam velocity. Investigation of power thresholds for curved ferromagnetic skyrmion motion however is outside the scope of the current work, and the results shown in Figure 5 demonstrate the possibility to guide skyrmions in any direction on a magnetic surface.

## V. Effect on Skyrmion Lattices

It is also possible to address multiple skyrmions using a single laser spot, and this is discussed here for a skyrmion lattice. Moreover, for experimental verification of the all-optical control mechanism presented here, it is advantageous to first investigate a simple case where the beam is stationary. A suggested case is shown in Figure 6, where a simple 2 μm × 2 μm square is used, hosting a ferromagnetic skyrmion lattice configuration. A stationary laser spot is centred on the sample, and computed skyrmion paths are plotted in Figure 6, with the initial skyrmion lattice configuration shown. The same forces discussed in previous sections are present, however we also have additional repulsive interactions between skyrmions. Thus the central skyrmion is stabilized at the laser spot centre, however the neighbouring skyrmions undergo a clockwise orbiting motion due to the Magnus effect ($Q = -1$). Skyrmions are attracted to the centre due to the combination of $F_{TG}$ and $F_{ME}$, with the effect decreasing away from the laser spot. Since the orbiting skyrmions are prevented from spiralling all the way towards the laser spot centre, owing to repulsive interactions with skyrmions closer or at the centre, the net effect is an orbiting motion for the entire lattice. Thus an experimental verification here could consist of observing the rotation of a skyrmion lattice using a suitable direct imaging technique, before and after application of a focused laser pulse of sufficiently large power and small spot-size.

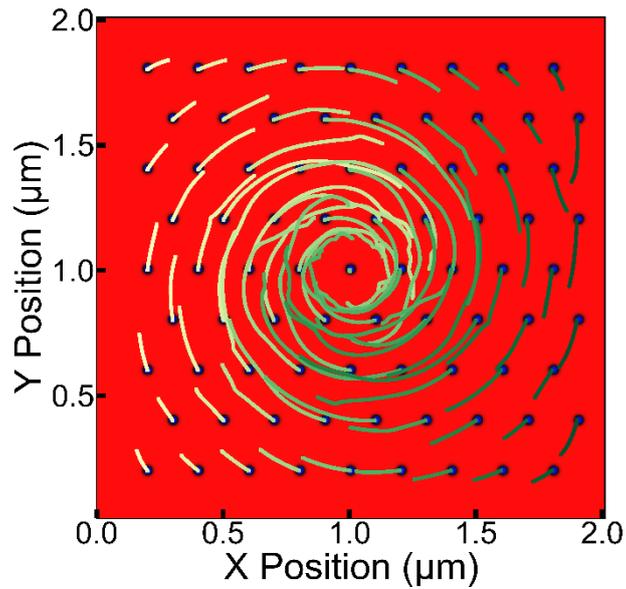

**Figure 6** – Effect of a static laser beam on a skyrmion lattice (skyrmions with $Q = -1$). Initial positions are shown, with overlaid computed paths in response to a centred laser spot with 1.5 mW power. Skyrmions rotate clockwise due to the Magnus force, with effect increasing closer to the centre; the skyrmion at the centre provides a repulsive force on neighbouring skyrmions.

A further example is given in the Supplemental Material, where the effect of a moving laser spot on a skyrmion lattice is shown. Due to the incomplete orbiting motion as the laser spot moves, a symmetry breaking effect is generated, where a net accumulation of skyrmions in the top or bottom halves of the sample is obtained depending on the beam travel direction. Further work is required to fully investigate the effect of magneto-thermo-elastic coupling on skyrmion lattices, and it is hoped the present work will stimulate efforts in this direction.

# VI. Conclusions

For magnetic thin films hosting skyrmions, using a focused laser beam to produce a local temperature gradient, and hence a thermo-elastic lattice expansion, it was shown that forces are generated due to magneto-thermo-elastic coupling. One such force is due to effective anisotropy gradients generated by local temperature gradients, acting towards the centre of the laser spot. Another equally important force however, is due to the ME effect as a result of thermo-elastic lattice expansion, and it was shown that for positive ME coupling coefficients this force also acts towards the centre of the laser spot, with the two effects reinforcing each other owing to increase in skyrmion diameter. Conversely, for negative ME coupling coefficients, the two forces nearly cancel out, largely due to their opposing directions. By moving the laser spot, ferromagnetic and antiferromagnetic skyrmion displacement was shown, with a threshold power dependent on the beam velocity. This method of all-optically generated skyrmion displacement allows full control of motion over two-dimensional magnetic surfaces, with a suggested method of MEMS-VCSEL on-chip integration, in contrast to current-induced skyrmion motion, or even SAW-generated skyrmion motion, which are limited by lithographically defined electrodes. It is hoped this work will stimulate future efforts in this direction, with methods of experimental verification also discussed, particularly for skyrmion lattices. Further possible extensions include use of synthetic antiferromagnetic multilayers, where the skyrmion Hall effect vanishes [62], but also combined current and focused laser pulse control to achieve skyrmion motion gating and controlled deflection.

## Appendix A

The magneto-thermo-elastic model has been implemented in BORIS [63] using finite differences, running on CUDA-enabled graphical processing units. The strain is related to mechanical displacement, **u**, as:

$$\varepsilon_{pq} = \frac{1}{2}\left(\frac{\partial u_p}{\partial q} + \frac{\partial u_q}{\partial p}\right), \quad (p,q = x, y, z) \tag{5}$$

Taking the time derivative of Equations (3) and (5) allows replacing the displacement with velocity, and together with the third expression in Equation (1) we obtain a system of first-order differential equations for elastodynamics, written in the velocity-stress representation as:

$$\rho \frac{\partial v_p}{\partial t} = \sum_{q=x,y,z} \frac{\partial \sigma_{pq}}{\partial q} - \eta v_p, \quad (p = x, y, z)$$

$$\frac{\partial \sigma_{pp}}{\partial t} = c_{11}\frac{\partial v_p}{\partial p} + c_{12}\left(\frac{\partial v_q}{\partial q} + \frac{\partial v_r}{\partial r}\right) - (c_{11} + 2c_{12})\alpha_T \frac{\partial T}{\partial t} - 2B_1 m_p \frac{\partial m_p}{\partial t}, \quad (p,q,r = x, y, z, p \neq q \neq r) \tag{6}$$

$$\frac{\partial \sigma_{pq}}{\partial t} = \frac{c_{44}}{2}\left(\frac{\partial v_p}{\partial q} + \frac{\partial v_q}{\partial p}\right) - 2B_2\left(m_p \frac{\partial m_q}{\partial t} + m_q \frac{\partial m_p}{\partial t}\right), \quad (p,q = x, y, z, p \neq q)$$

Time derivatives of temperature and magnetization, necessary for Equation (6), are evaluated directly in the dynamical magneto-thermo-elastic model of Equation (1).

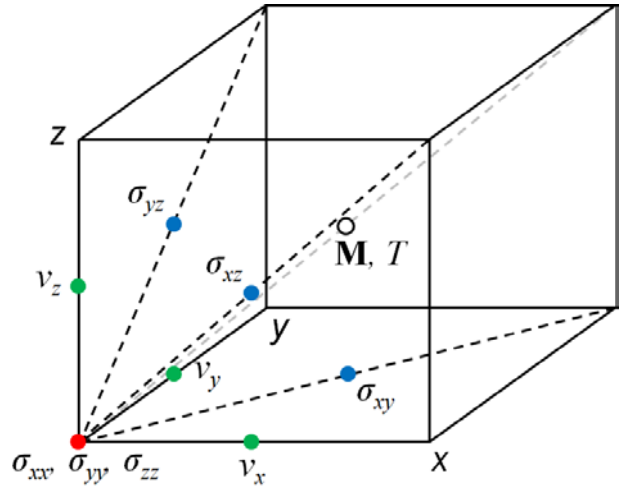

**Figure 7** – Velocity, stress, magnetization and temperature components for finite difference discretization of the magneto-thermo-elastic model. Velocity components are edge-centred, diagonal stress components are at vertices, shear stress components are face-centred, and magnetization and temperature values are cell-centred.

The equations are solved using the FDTD scheme, with staggered velocity and stress components as indicated in Figure 7 for a given computational cell. The temperature and magnetization values are cell-centred; their values and that of their time derivative at vertices, edge and face centres, are obtained to second order accuracy in space using interpolation. Spatial derivatives at inner points are evaluated using standard finite differences to second order accuracy [64]. There must be at least one fixed surface where the mechanical displacement is zero, since no translational motion is allowed, and thus the velocity is also zero. Velocity components on a fixed surface are set directly as zero, whilst derivatives of velocity components perpendicular to a fixed surface are evaluated using standard expressions with Dirichlet boundary condition of zero [64]. As an example, for a fixed $xz$ surface in Figure 7, $v_x = v_z = 0$, whilst for $\partial v_y / \partial y$ at the boundary the Dirichlet condition $v_y = 0$ applies. For all other surfaces (free surfaces) boundary stress values are prescribed from external forces as ($\delta$ is the Kronecker delta):

$$\sum_{q=x,y,z} \sigma_{pq} \delta_{rq} = F_r, \quad (p, r = x, y, z) \tag{7}$$

For Equation (7) $p$ is normal to the free surface. In this work there are no external forces, thus **F** is zero on all surfaces. The boundary stress values are either set directly for components on free surfaces, or else are used as Dirichlet boundary conditions to evaluate derivatives of stress perpendicular to a free surface. As an example, for a fixed $xz$ surface in Figure 7, $\sigma_{yy} = F_y$ is set directly, whilst $\sigma_{xy} = F_x$ and $\sigma_{yz} = F_z$ are used as Dirichlet boundary conditions to evaluate $\partial \sigma_{xy} / \partial y$ and $\partial \sigma_{yz} / \partial y$ respectively, at the boundary. Finally, required derivatives of velocity perpendicular to a free surface are obtained from the second expression in Equation (6), using boundary values for the time derivative of stress, in turn obtained from the time derivative of Equation (7) (also zero in this work). These are obtained in general as ($p$ is normal to the free surface):

$$\frac{\partial v_p}{\partial p} = \frac{1}{c_{11}} \frac{\partial F_p}{\partial t} - \frac{c_{12}}{c_{11}} \left( \frac{\partial v_q}{\partial q} + \frac{\partial v_r}{\partial r} \right) + \left( 1 + 2 \frac{c_{12}}{c_{11}} \right) \alpha_T \frac{\partial T}{\partial t} + \frac{2 B_1}{c_{11}} m_p \frac{\partial m_p}{\partial t}, \quad (p, q, r = x, y, z, p \neq q \neq r) \tag{8}$$

For multi-layered structures, such as the FM/HM bilayer investigated here, composite media boundary (CMB) conditions are also required. Using the scheme in Figure 7, these consist of enforcing continuity of velocity and stress components located on the CMB interface, achieved by interpolating respective values either side of the interface. These values are set after the elastodynamics equation is iterated in all layers. It is also necessary to specify initial values for stress and velocity. The velocity initial value is zero throughout, whilst the initial values of stress components, which are not on a free surface where Equation (7) applies, are computed using Equation (3) by taking the initial strain to be zero.

The LLG equation is solved using the implemented RK4 method [65] with 200 fs time-step, using the temperature and strain values available at the start of each time step. The heat equation is solved using the finite time centred-space method with 100 fs time-step. The elastodynamics equation is solved with velocity and stress staggered in time, i.e. velocity is updated first, and resulting values are used to update the stress using Equation (6). The time-step for the elastodynamics equation is $\Delta t = 100$ fs, which satisfies the Courant, Friedrichs, Lewy condition, $\Delta t < 1/\left(v_\text{p}\sqrt{\Delta x^{-2} + \Delta y^{-2} + \Delta z^{-2}}\right)$, where $v_\text{p}$ is the elastic compressional wave velocity ($v_\text{p} = 6000$ m/s here), and $\Delta x$, $\Delta y$, $\Delta z$ are the cell-size dimensions. The FM layer is discretized as (4 nm, 4 nm, 1 nm) for the elastodynamics and heat equations, whilst the magnetic cell-size is 2 nm. The HM layer is discretized as (4 nm, 4 nm, 5 nm) for the elastodynamics and heat equations. Finally, material values not specified in the main text are $C = 420$ J/kgK, $\rho = 8920$ kg/m$^3$, $K = 122$ W/mK for FM (bulk Co values), and $C = 125.6$ J/kgK, $\rho = 21452$ kg/m$^3$, $K = 71.6$ W/mK for HM (bulk Pt values). For both FM and HM the mechanical damping is $\eta = 10^{15}$ kg/m$^3$s, although this does not affect strain values obtained on the time-scale of magnetization processes, and serves to absorb the much faster transient elastic waves.

# All-optical magneto-thermo-elastic skyrmion motion

Serban Lepadatu[1]

[1]*Jeremiah Horrocks Institute for Mathematics, Physics and Astronomy, University of Central Lancashire, Preston PR1 2HE, U.K.*

Supplemental Material

## I. Effect of Magneto-Elastic Coupling Sign

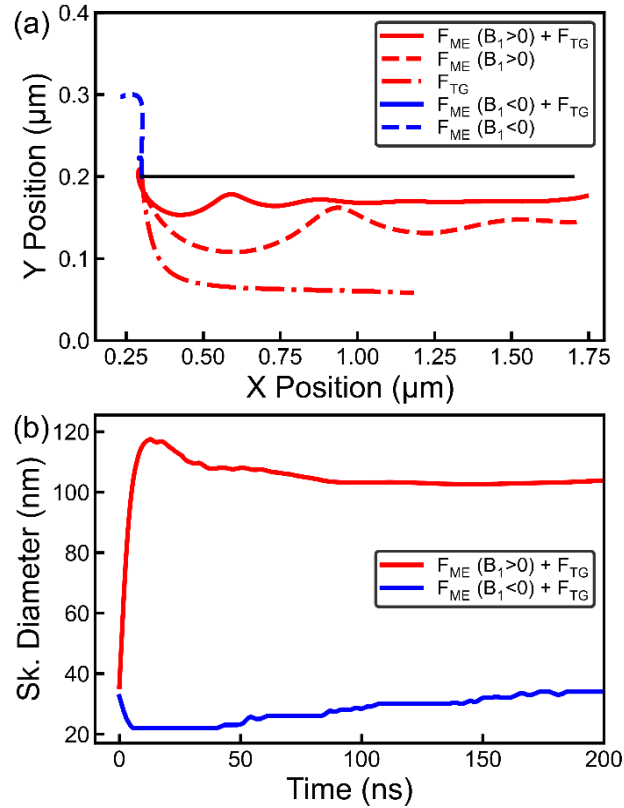

**Figure S1** – Effect of ME coupling sign on skyrmion motion ($Q = +1$) for 1 mW absorbed power, with laser spot scanned at 5 m/s through the centre of a track 400 nm wide and 2 μm long. (a) The black line shows the laser scanning path, starting at $x = 0.3$ μm, $y = 0.2$ μm. The red lines are reproduced from Figure 2 for positive ME coupling ($B_1 = 10$ MJ/m$^3$), where the force $F_{ME}$ points towards the laser spot centre. The blue lines show the resultant skyrmion paths for negative ME coupling ($B_1 = -10$ MJ/m$^3$), where $F_{ME}$ points away from the laser spot centre, opposing $F_{TG}$ (the dashed blue line shows the effect of $F_{ME}$ only, whilst the solid line shows the combined effect of $F_{ME}$ and $F_{TG}$). (b) Skyrmion diameter as a function of time for different ME coupling signs.

As discussed in the main text, for positive ME coupling $F_{ME}$ acts towards the centre of the laser spot, and thus in the same direction as $F_{TG}$. Resultant skyrmion paths under the action of separate forces, as well as combination, are plotted again in Figure S1(a). If $B_1$ is negative the sign of $F_{ME}$ is reversed, and thus opposes $F_{TG}$. This is also shown in Figure S1(a), where the dashed blue line shows the effect of $F_{ME}$ alone, whilst the solid blue line shows the combined effect. The combination of $F_{ME}$ and $F_{TG}$ nearly cancels out due to opposite directions on the one hand, with the effect on skyrmion motion also reduced due to a significant reduction in skyrmion diameter, as shown in Figure S1(b). For negative ME coupling a tensile strain induces an easy axis, and a hard plane perpendicular to the strain axis, and thus the increase in total out-of-plane anisotropy results in a decrease in the skyrmion diameter. Moreover, the blue line paths are considerably shorter since the laser spot and skyrmion are moving away from each other.

## II. Absorbed Power Dependences

As the laser absorbed power increases, the maximum temperature, and thus the perpendicular strain due to the thermo-elastic effect, also increase as expected. This is shown Figure S2(a), with increases in temperature and strain depending linearly on the absorbed power. Increasing the power decreases the effective magnetic anisotropy, $K_{eff}(T) = m_e^2(K_U m_e - \mu_0 M_S^2/2)$, with the resulting gradient in $K_{eff}$ resulting in the $F_{TG}$ force. The maximum gradient in $K_{eff}$ is plotted in Figure S2(b) as a function of power. Since for positive ME coupling a hard axis is induced along the tensile strain, an anisotropy gradient in the ME anisotropy energy density, $E_{ME}$, is also obtained, resulting in the $F_{ME}$ force. The maximum gradient in $E_{ME}$ is also plotted in Figure S2(b). Finally, due to decrease in the total anisotropy with temperature, the skyrmion diameter increases with power. This is plotted in Figure S2(c) both for a ferromagnetic ($Q = 1$), and antiferromagnetic skyrmion ($Q = 0$), obtained during a scan with a 10 m/s laser spot velocity, plotted up to the instability threshold. The diameter dependence on power is reduced for $Q = 0$ since the demagnetizing field contribution is negligible. Thus the instability threshold (boundary between regions I and IIa – see main text) is increased for antiferromagnetic skyrmions.

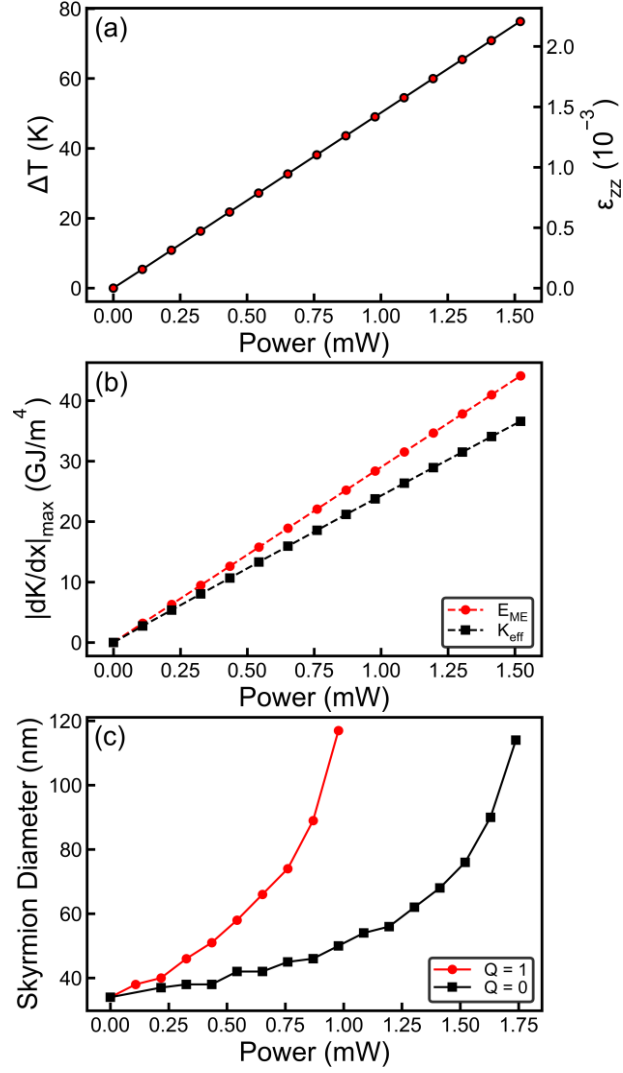

**Figure S2** – Variation of important parameters with laser power for a static laser beam in the centre of the 400 nm wide and 2 μm long track. (a) Maximum temperature change and maximum perpendicular strain, reached at the center of the laser spot. (b) Maximum anisotropy gradients, shown for the effective anisotropy, $K_{eff} = K_U - \mu_0 M_S^2/2$, and the strain-induced anisotropy energy density, $E_{ME}$. (c) Skyrmion diameter as a function of laser power for a ferromagnetic ($Q = 1$), and antiferromagnetic skyrmion ($Q = 0$), obtained during a scan with a 10 m/s laser spot velocity, plotted up to the instability threshold.

## III. Effect on Skyrmion Lattices

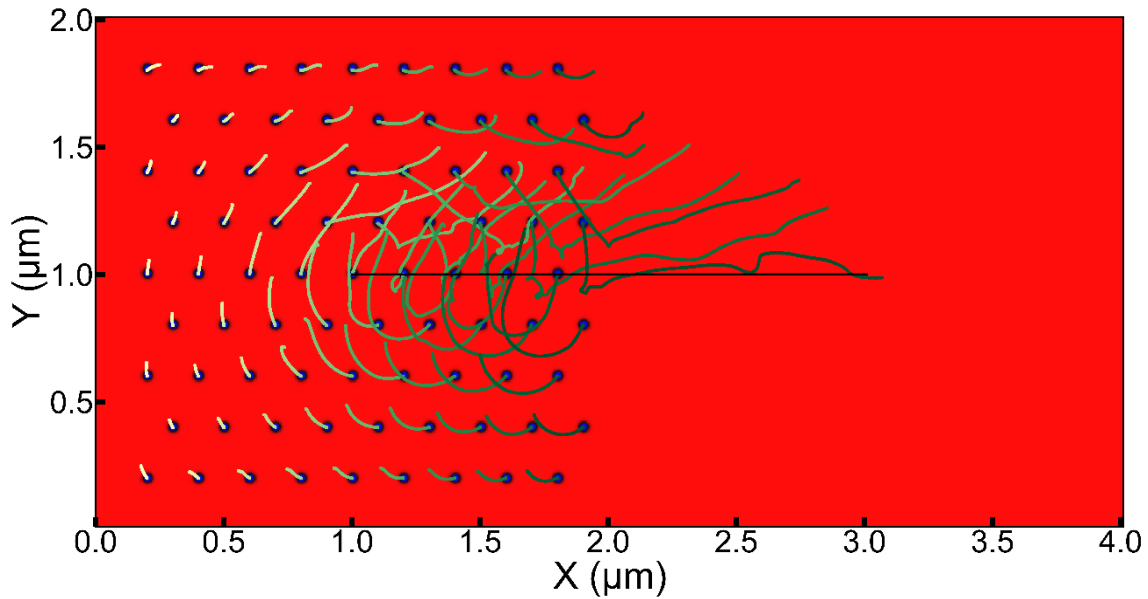

**Figure S3** – Effect of a laser spot with velocity of 10 m/s and 1.5 mW absorbed power on a skyrmion lattice (skyrmions with $Q = -1$). The laser path is indicated using the solid black line. Initial skyrmion positions are shown, with overlaid computed paths.

In the main text the effect of a static laser spot on a skyrmion lattice was discussed. Here the effect of a moving laser spot on a skyrmion lattice is also shown. In Figure S3 a 4 μm × 2 μm rectangle is used for the FM/HM bilayer, hosting a skyrmion lattice in the left half. A laser spot with 10 m/s velocity originates at $x = 1$ μm, $y = 1$ μm, terminating at $x = 3$ μm, $y = 1$ μm, and with absorbed power of 1.5 mW. As for the case with a static laser spot, skyrmions tend to rotate clockwise due to the Magnus force, however skyrmions in the top-right part acquire an overall anti-clockwise rotation due to repulsive interactions with skyrmions originating from the lower half. The effect is a net accumulation of skyrmions in the top half. Thus, unlike the static laser spot case, due to the incomplete orbiting motion as the laser spot moves, a symmetry breaking effect is generated, where a net accumulation of skyrmions is obtained in the top half. Reversing the beam travel direction results in a net accumulation of skyrmions in the bottom half.

## IV. Skyrmion Deformation

A threshold power is required for skyrmion motion, depending on the beam velocity, as discussed in the main text. At higher laser powers however, the skyrmion becomes unstable and can be deformed into a skyrmion string. This process is shown in Figure S4. As the skyrmion diameter increases due to reduction in total anisotropy, it becomes elongated, and is distorted under the uneven action of $F_{ME}$, $F_{TG}$ and Magnus forces on its elongated boundary. The skyrmion string also moves with the laser spot due to attractive forces towards the centre.

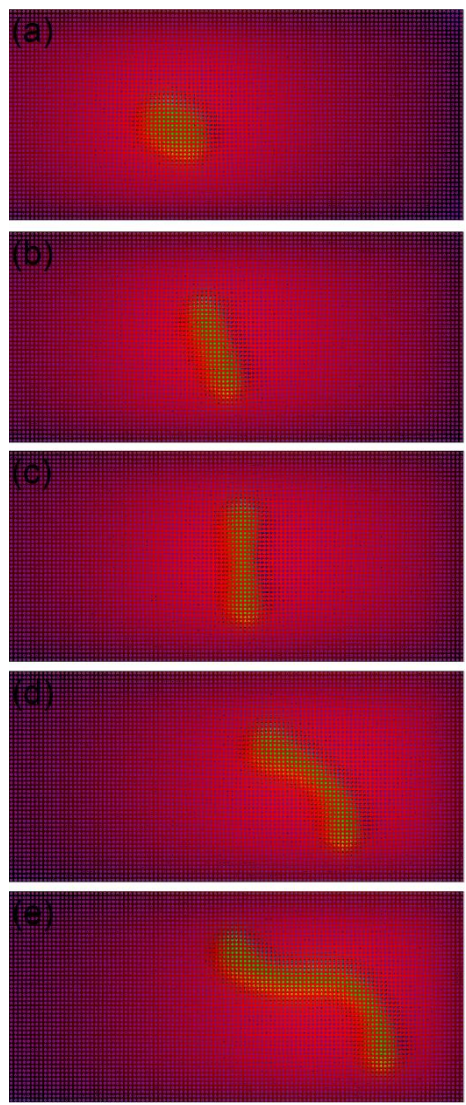

**Figure S4** – Deformation of a skyrmion into a skyrmion string during motion induced by a laser spot with 5 m/s velocity and 1.1 mW absorbed power in the FM/HM bilayer track. The images show the skyrmion overlaid over the color-coded $z$-direction strain for illustration (red indicates maximum strain at the centre of the laser spot). (a)-(e) Snapshots of the deformation process at 10 ns intervals.

5